\documentstyle[12pt
]{article}

\catcode`\@=11

\def\ps@myheadings{\let\@mkboth\@gobbletwo
\def\@oddhead{\hbox{} 
\rightmark\hfil\ninerm\thepage}   
\def\@oddfoot{}\def\@evenhead{\ninerm\thepage\hfil 
\leftmark\hbox{}}\def\@evenfoot{}
\def\sectionmark##1{}\def\subsectionmark##1{}}

\begin{document}

\newcommand{\symbolfootnote}{\renewcommand{\thefootnote}
        {\fnsymbol{footnote}}}
\renewcommand{\thefootnote}{\fnsymbol{footnote}}
\newcommand{\alphfootnote}
        {\setcounter{footnote}{0}
         \renewcommand{\thefootnote}{\sevenrm\alph{footnote}}}

\newcounter{sectionc}\newcounter{subsectionc}\newcounter{subsubsectionc}
\renewcommand{\section}[1] {\vspace{0.6cm}\addtocounter{sectionc}{1} 
\setcounter{subsectionc}{0}\setcounter{subsubsectionc}{0}\noindent 
        {\bf\thesectionc. #1}\par\vspace{0.4cm}}
\renewcommand{\subsection}[1] {\vspace{0.6cm}\addtocounter{subsectionc}{1} 
        \setcounter{subsubsectionc}{0}\noindent 
        {\it\thesectionc.\thesubsectionc. #1}\par\vspace{0.4cm}}
\renewcommand{\subsubsection}[1] {\vspace{0.6cm}\addtocounter{subsubsectionc}{1}
        \noindent {\rm\thesectionc.\thesubsectionc.\thesubsubsectionc. 
        #1}\par\vspace{0.4cm}}
\newcommand{\nonumsection}[1] {\vspace{0.6cm}\noindent{\bf #1}
        \par\vspace{0.4cm}}
                                                 
\newcounter{appendixc}
\newcounter{subappendixc}[appendixc]
\newcounter{subsubappendixc}[subappendixc]
\renewcommand{\thesubappendixc}{\Alph{appendixc}.\arabic{subappendixc}}
\renewcommand{\thesubsubappendixc}
        {\Alph{appendixc}.\arabic{subappendixc}.\arabic{subsubappendixc}}

\renewcommand{\appendix}[1] {\vspace{0.6cm}
        \refstepcounter{appendixc}
        \setcounter{figure}{0}
        \setcounter{table}{0}
        \setcounter{equation}{0}
        \renewcommand{\thefigure}{\Alph{appendixc}.\arabic{figure}}
        \renewcommand{\thetable}{\Alph{appendixc}.\arabic{table}}
        \renewcommand{\theappendixc}{\Alph{appendixc}}
        \renewcommand{\theequation}{\Alph{appendixc}.\arabic{equation}}
        \noindent{\bf Appendix \theappendixc #1}\par\vspace{0.4cm}}
\newcommand{\subappendix}[1] {\vspace{0.6cm}
        \refstepcounter{subappendixc}
        \noindent{\bf Appendix \thesubappendixc. #1}\par\vspace{0.4cm}}
\newcommand{\subsubappendix}[1] {\vspace{0.6cm}
        \refstepcounter{subsubappendixc}
        \noindent{\it Appendix \thesubsubappendixc. #1}
        \par\vspace{0.4cm}}

\def\abstracts#1{{
        \centering{\begin{minipage}{30pc}\tenrm\baselineskip=12pt\noindent
        \centerline{\tenrm ABSTRACT}\vspace{0.3cm}
        \parindent=0pt #1
        \end{minipage} }\par}} 

\newcommand{\bibit}{\it}
\newcommand{\bibbf}{\bf}
\renewenvironment{thebibliography}[1]
        {\begin{list}{\arabic{enumi}.}
        {\usecounter{enumi}\setlength{\parsep}{0pt}
\setlength{\leftmargin 1.25cm}{\rightmargin 0pt}
         \setlength{\itemsep}{0pt} \settowidth
        {\labelwidth}{#1.}\sloppy}}{\end{list}}

\topsep=0in\parsep=0in\itemsep=0in
\parindent=1.5pc

\newcounter{itemlistc}
\newcounter{romanlistc}
\newcounter{alphlistc}
\newcounter{arabiclistc}
\newenvironment{itemlist}
        {\setcounter{itemlistc}{0}
         \begin{list}{$\bullet$}
        {\usecounter{itemlistc}
         \setlength{\parsep}{0pt}
         \setlength{\itemsep}{0pt}}}{\end{list}}

\newenvironment{romanlist}
        {\setcounter{romanlistc}{0}
         \begin{list}{$($\roman{romanlistc}$)$}
        {\usecounter{romanlistc}
         \setlength{\parsep}{0pt}
         \setlength{\itemsep}{0pt}}}{\end{list}}

\newenvironment{alphlist}
        {\setcounter{alphlistc}{0}
         \begin{list}{$($\alph{alphlistc}$)$}
        {\usecounter{alphlistc}
         \setlength{\parsep}{0pt}
         \setlength{\itemsep}{0pt}}}{\end{list}}

\newenvironment{arabiclist}
        {\setcounter{arabiclistc}{0}
         \begin{list}{\arabic{arabiclistc}}
        {\usecounter{arabiclistc}
         \setlength{\parsep}{0pt}
         \setlength{\itemsep}{0pt}}}{\end{list}}

\newcommand{\fcaption}[1]{
        \refstepcounter{figure}
        \setbox\@tempboxa = \hbox{\tenrm Fig.~\thefigure. #1}
        \ifdim \wd\@tempboxa > 6in
           {\begin{center}
        \parbox{6in}{\tenrm\baselineskip=12pt Fig.~\thefigure. #1 }
            \end{center}}
        \else
             {\begin{center}
             {\tenrm Fig.~\thefigure. #1}
              \end{center}}
        \fi}

\newcommand{\tcaption}[1]{
        \refstepcounter{table}
        \setbox\@tempboxa = \hbox{\tenrm Table~\thetable. #1}
        \ifdim \wd\@tempboxa > 6in
           {\begin{center}
        \parbox{6in}{\tenrm\baselineskip=12pt Table~\thetable. #1 }
            \end{center}}
        \else
             {\begin{center}
             {\tenrm Table~\thetable. #1}
              \end{center}}
        \fi}

\def\@citex[#1]#2{\if@filesw\immediate\write\@auxout
        {\string\citation{#2}}\fi
\def\@citea{}\@cite{\@for\@citeb:=#2\do
        {\@citea\def\@citea{,}\@ifundefined
        {b@\@citeb}{{\bf ?}\@warning
        {Citation `\@citeb' on page \thepage \space undefined}}
        {\csname b@\@citeb\endcsname}}}{#1}}

\newif\if@cghi
\def\cite{\@cghitrue\@ifnextchar [{\@tempswatrue
        \@citex}{\@tempswafalse\@citex[]}}
\def\citelow{\@cghifalse\@ifnextchar [{\@tempswatrue
        \@citex}{\@tempswafalse\@citex[]}}
\def\@cite#1#2{{$\null^{#1}$\if@tempswa\typeout
        {IJCGA warning: optional citation argument 
        ignored: `#2'} \fi}}
\newcommand{\citeup}{\cite}

\def\fnm#1{$^{\mbox{\scriptsize #1}}$}
\def\fnt#1#2{\footnotetext{\kern-.3em
        {$^{\mbox{\sevenrm #1}}$}{#2}}}

\font\twelvebf=cmbx10 scaled\magstep 1
\font\twelverm=cmr10 scaled\magstep 1
\font\twelveit=cmti10 scaled\magstep 1
\font\elevenbfit=cmbxti10 scaled\magstephalf
\font\elevenbf=cmbx10 scaled\magstephalf
\font\elevenrm=cmr10 scaled\magstephalf
\font\elevenit=cmti10 scaled\magstephalf
\font\bfit=cmbxti10
\font\tenbf=cmbx10
\font\tenrm=cmr10
\font\tenit=cmti10
\font\ninebf=cmbx9
\font\ninerm=cmr9
\font\nineit=cmti9
\font\eightbf=cmbx8
\font\eightrm=cmr8
\font\eightit=cmti8

\newcommand{\bea}{\begin{eqnarray}}
\newcommand{\beal}[1]{\begin{eqnarray}\label{#1}}
\newcommand{\eea}{\end{eqnarray}}
\newcommand{\beq}{\begin{equation}}
\newcommand{\eeq}{\end{equation}}
\newcommand{\bel}[1]{\begin{equation}\label{#1}}
\newcommand{\nn}{\nonumber}

\newcommand{\Sp}{\,\mbox{tr}\,}

\newcommand{\sumint}
  {{\textstyle\sum}\!\!\!\!\!\!\int\nolimits_{\;P}}
\newcommand{\sumints}
  {{\scriptstyle\Sigma}\!\!\!\!\int\nolimits_{P}}

\newcommand{\w}{\omega}
\renewcommand{\a}{\alpha}
\renewcommand{\b}{\beta}
\renewcommand{\d}{\delta}
\newcommand{\s}{\sigma}
\renewcommand{\L}{\Lambda}
\newcommand{\G}{\Gamma}
\newcommand{\g}{\gamma}
\newcommand{\bg}{{\mbox{\protect\boldmath $\gamma$}}}

\newcommand{\mn}{{\mu\nu}}

\newcommand{\cl}[1]{{\cal #1}}

\newcommand{\0}{\over }
\newcommand{\1}[1]{\frac{1}{#1}}
\newcommand{\2}{\frac{1}{2}}
\newcommand{\4}{\frac{1}{4}}
\newcommand{\6}{\partial}
\newcommand{\with}{\quad\mbox{with}\quad}
\newcommand{\quer}[1]{\overline{#1}}

\def\({\left(}     \def\){\right)}
\def\lek{\left[}   \def\rek{\right]}
\def\lgk{\left\{}  \def\rgk{\right\}}
\newcommand{\rang}{\right\rangle}
\newcommand{\lang}{\left\langle}

\def\lesssim{\mbox{\,\raisebox{.3ex}{
           $<$}$\!\!\!\!\!$\raisebox{-.9ex}{$\sim$}\,\,}}
\def\gtrsim{\mbox{\,\raisebox{.3ex}{
           $>$}$\!\!\!\!\!$\raisebox{-.9ex}{$\sim$}\,\,}}

\baselineskip=16pt
\begin{flushright}
\tt 
TUW 96-15\\
\end{flushright}
\vspace{0.4cm}
\centerline{{\tenbf HARD THERMAL LOOPS 
NEAR THE LIGHT-CONE}\footnote{Invited talk at the
Workshop on Quantum Chromodynamics, 3--8 June 1996, American
University of Paris, France.}
}
\vspace{0.8cm}
\centerline{\tenrm Anton REBHAN}
\baselineskip=13pt
\centerline{\tenit Inst. f. Theoret. Phys., Techn. Univ. Wien}
\baselineskip=12pt
\centerline{\tenit Wiedner Hauptstr. 8--10/136, A-1040 Vienna, Austria}
\vspace{0.9cm}
\abstracts{In hot gauge theories, perturbation theory at the
scale of the Debye screening mass requires the
resummation of the so-called hard thermal loops, which corresponds to
using an effective action obtained by integrating out the modes with
momentum of order of the temperature. As is well-known,
quantities which are sensitive to
the nonperturbative magnetic screening mass still remain incalculable
in this resummed perturbation theory. A different
breakdown of the latter occurs whenever external momenta are light-like,
because the hard thermal loops themselves develop collinear singularities.
However, taking into account asymptotic thermal masses for
the hard modes regulates the hard thermal loops without spoiling their
gauge invariance. 
}

\vspace{0.8cm}
\twelverm   
\baselineskip=14pt

\section{Introduction}

At the scale of the order of the Debye and plasmon masses, hot gauge
theories\cite{appl} 
are no longer adequately described by ordinary thermal perturbation
theory. A consistent perturbative approach requires (at least) the
resummation of the so-called hard thermal loops (HTL)\cite{BP}. 
The latter can be
summarized by compact gauge-invariant effective actions which have
been found by Taylor and Wong, and by Braaten and Pisarski 
\cite{eff,shortcut,BPeff,Nair}, to wit
\beq
\cl S_{\mathrm eff} = -{3m^2\0 4} \int\! d^4x F_a^{\mu\a}(x)
 \lang{Y_\a Y_{\beta} \0 (YD)^2_{ab}}\rang
 F_b^\beta{}_\mu(x) 
-M_f^2\int\! d^4x \quer\psi(x)
  \lang{Y^\mu\0 iYD}\rang
  \g_\mu\psi(x) \;
\label{effBP}
\eeq
where $m$ and $M_f$ are the long-wavelength plasma masses of
the gauge bosons and the fermions, resp., and $Y=(1,\vec e)$ is
a normalized light-like vector whose spatial direction is
averaged over in $\lang\ldots\rang$.

This approach has been applied successfully to a number of 
problems\cite{appl}, and even when the quantities under consideration are
sensitive to the magnetic mass scale, it is usually possible to
extract a leading logarithmic correction. A quite different shortcoming
of the by now standard HTL resummation has been encountered recently
in the attempt to calculate the production rate of soft real photons
from a quark-gluon-plasma\cite{BPS}. This requires to use hard
thermal loops with external momenta on the light-cone, for which
collinear singularities appear.

In a somewhat different context, similar difficulties can be observed already
in the simple case of hot scalar electrodynamics\cite{KRS}.

\section{Longitudinal plasmons in scalar electrodynamics}

In hot scalar electrodynamics, the matter part of the
HTL effective action (\ref{effBP}) is replaced by a simple
mass term,
\beq
\cl S_{\mathrm eff} =  -m_{\mathrm sc.}^2 \int\! d^4x |\phi|^2,
\qquad m_{\mathrm sc.}={eT\02}
\label{mscalar}
\eeq
which makes it possible to calculate analytically
the complete next-to-leading order corrections to the
dispersion law of the photonic plasma excitations. It turns
out that for both, the transverse and the longitudinal branches,
the frequencies $\omega_{t,\ell}(q)$ associated with a given wave-vector
$\vec q$ are diminished by corrections $\sim e$. However, in the
case of the longitudinal plasmons, the dispersion curve approaches
the light-cone exponentially at leading order, and the negative
correction appears to make this curve pierce through the light-cone.
While there is nothing wrong with that (the group velocity
remains smaller than the speed of light), it certainly constitutes
a qualitative change in the dispersion laws
which a merely perturbative treatment never can
decide.

Indeed, by inspecting the behaviour of $\Pi_{00}(Q)$ as $Q^2\to0$,
one finds that the correction term goes like
\beq
\delta \Pi_{00}(Q) \sim {em^2q\0\sqrt{Q^2}}
\label{dPi}
\eeq
whereas the HTL piece is only logarithmically singular. So for
$Q^2/q^2 \ll e^2$, $\delta \Pi_{00}\gg \Pi^{\mathrm HTL}_{00}$,
and perturbation theory clearly cannot be trusted any longer.

The difficulty is in fact rooted already in the singular behaviour
of the HTL part of $\Pi_{00}$.  Universally (i.e. up to different
normalizations in different theories), it reads \cite{KalKlW}
\bea
\Pi_{00}(Q) &=& 4e^2 \int\!{d^3p \0 (2\pi)^3}\,n(p)
\lgk 1-{Q_0 \0
 Q_0-\vec p\vec q/p} \rgk \nn\\
&=& 3m^2  \(1 - {Q_0 \0 2q} \ln{Q_0+q\0Q_0-q}\) \;.
\label{poohtl}\eea
In this form,
the logarithmic singularity for light-like momenta is seen to
come from a collinear singularity appearing when the hard
momentum $\vec p$ becomes parallel to the soft $\vec q$.
Any mass for the hard modes would cut off this singularity.

In fact, the scalar particles in the plasma always have
a thermal mass associated with them, for the latter is
a constant, independent of momentum, cf.\ (\ref{mscalar}).
Including this thermal mass of the scalars already at the
HTL level indeed removes the behaviour (\ref{dPi}), and
modifies the leading order piece such that it is finite up to
and including the light-cone. This gives a finite value of
$|\vec q|$, where the longitudinal plasmon dispersion curve
hits the light-cone,
\beq
q_c^2=-\Pi_{00}(Q_0=q=q_c)={e^2T^2\03}
\(\ln{4\0e}+{1\02}-\g+{\zeta'(2)\0\zeta(2)}+O(e)\)
\eeq
where $\ln(4/e)$ comes from
$\ln(2T/m_{\mathrm sc.})$.

There is still a logarithmic branch cut for $|Q_0|<q$ and a large
imaginary part for space-like momenta, which overdamps any
excitations far from the light-cone. 
However, since the singularity at the light-cone
has gone, this imaginary part sets in smoothly\cite{KRS}, so that there is a
small but finite range just across the light-cone
in the space-like region with weakly damped plasmons.\cite{SULS}

\section{Improved hard thermal loops for QED and QCD}

In spinor QED and QCD, 
the dispersion laws are much more complicated than the one of hot
scalars with the simple mass term
(\ref{mscalar}). There are branches corresponding to collective modes
that have no analogue in the vacuum theory---the longitudinal plasmons
and the fermionic plasminos. These have effective thermal masses
which approach zero as the momentum increases, but at the same time
the residues of the corresponding poles in the propagator go to zero as well,
and they do so exponentially fast\cite{Pispha}. On the other hand, the branches
connecting to the ordinary degrees of freedom retain nonvanishing
thermal masses, which approach a constant value and read
\beq
m_\infty^2={3\02}m^2,\quad
M_{\infty}^2= 2 M_f^2
\eeq
for gauge bosons and fermions, respectively.

These asymptotic thermal masses are generated by the self-interactions
of the {\em hard} modes among themselves. In the effective (HTL) action
obtained by integrating out the hard modes, these asymptotic masses
are negligible everywhere except close to the light-cone. There they
enter as a cutoff to the collinear singularities of the conventional
hard thermal loops.

While there was obviously no problem with gauge invariance with
the constant mass term in hot scalar electrodynamics, one might
expect problems with gauge invariance through the inclusion of 
an asymptotic thermal mass for photons and gluons.
However, it turns out that the improved HTL effective action which is
regular at the light-cone is still gauge invariant\cite{FR}.
In fact, the compact expressions of (\ref{effBP}) can be taken
over with such modifications that manifest gauge invariance is retained.
In the gluonic part these modifications merely amount to redefining
the light-like vector $Y$ into a time-like one and extending the
averaging prescription to a certain integral over $Y_0$. In the fermionic
part, there is instead a slight change in the functional form, but which
does not spoil manifest gauge invariance either.

The gauge invariance of the improved effective action is certainly
encouraging  for setting up a likewise improved resummed perturbation
theory. However, the question whether it is already sufficient for
having a well-behaved perturbation series in the vicinity of the light-cone
is still open.

Another candidate for the removal of the collinear singularities would
be the inclusion of damping of the hard modes. In the process of integrating
out the hard modes, this should be negligible compared to the effect of
thermal masses because the damping
produced by hard self-interactions is $\g\sim g^4T$. However,
through the interactions
with {\em soft} modes, there is an anomalously large damping $\g$ for the
hard modes of the order $g^2T\ln(1/g)$.\cite{LS,damping}
Although formally still of higher order,
such a damping would be even more effective in cutting off the collinear
singularities than the asymptotic mass, since (roughly)
\beq
{Q^2\0q^2}\to {Q_0\g\0q^2}\sim g \gg {m_\infty^2\0T^2} \sim g^2.
\eeq
(With $\g\sim g^4T$ the asymptotic thermal mass term would have won.)

However, in contrast to a resummation of asymptotic thermal masses,
it is generally not sufficient to modify only the propagators, when
damping is to be taken into account. Including damping only into the
propagators violates the Ward identities. In Abelian theories the latter
imply $\Pi_{\mu\nu}Q^\nu\equiv 0$, but one finds e.g.
\beq
\Pi_{0\nu}(Q_0,0)Q^\nu\Big|_{\vec q=0}={2i\g Q_0\0Q_0+2i\g}{e^2T^2\03}.
\eeq
Correcting also the vertices by sort of a ladder resummation
restores gauge invariance, but it turns out that $\g$ gets eliminated
not only from $\Pi_{\mu\nu}Q^\nu$, but from all the components of
$\Pi_{\mu\nu}$ !\cite{LS,KRS}

If this holds true more generally, then one may expect that the resummation
of the asymptotic thermal masses is indeed the relevant mechanism to
screen the collinear singularities. 

On the other hand,
there is recent work on the soft-real-photon production 
rate\cite{AGKP} which seems to indicate 
that certain formally higher-order contributions
dominate over lower-order when regularizing the light-cone singularities
by resumming the asymptotic thermal masses. 
By the same token,
we have found similar difficulties 
in the imaginary part of the
plasmon polarization tensor at one-loop-resummed order.
We expect that a careful study of the light-cone properties of the
resummed plasmon polarization tensor will elucidate the question of
how one has to improve thermal perturbation theory when light-like external
momenta are involved\cite{FRprep}.

\vskip 14pt

This work has been done in collaboration with F. Flechsig and H. Schulz
(Univ.\ Hannover), and U. Kraemmer (TU Wien). 
It has been
supported in part by the Austrian 
``Fonds zur F\"orderung der wissenschaftlichen
Forschung (FWF)'', project no. P10063-PHY and by
the EEC Program ``Human Capital
and Mobility'', Network ``Physics at High Energy Colliders'', contract
CHRX-CT93-0357 (DG 12 COMA). Moreover,
I would like to thank 
R. Baier, E. Petitgirard, S. Peign\'e, and R. Pisarski for discussions,
as well as
H. Fried, B. M\"uller, and
M. LeBellac for organizing this stimulating workshop.

\vspace{0.8cm}
\leftline{\twelvebf References}

\end{document}